\documentstyle[prd,preprint,aps,epsfig,floats]{revtex}

\begin{document}

\tightenlines
\input epsf.tex
\def\DESepsf(#1 width #2){\epsfxsize=#2 \epsfbox{#1}}
\draft
\thispagestyle{empty}
\preprint{\vbox{ \hbox{UMD-PP-02-014}
\hbox{October 2001}}}

\title{\Large \bf Big Bang Nucleosynthesis Constraints on Bulk Neutrinos}

\author{\large H. S. Goh\footnote{e-mail: hsgoh@glue.umd.edu}
 and R.N. Mohapatra\footnote{e-mail:rmohapat@physics.umd.edu}}

\address{ Department of Physics, University of Maryland\\
College Park, MD 20742, USA}

\maketitle

\thispagestyle{empty}

\begin{abstract}
We examine the constraints imposed by the requirement of successful
nucleosynthesis on models with one large extra hidden space dimension and
a single bulk neutrino.  We first use
naive  out  of equilibrium conditions to constrain the size of the extra
dimensions and the mixing between the active and the bulk neutrino. 
We then use the solution of the Boltzman kinetic equation
for the thermal distribution of the Kaluza-Klein modes and evaluate their
contribution to the energy density at the BBN epoch to constrain
the same parameters.
\end{abstract}

\section{Introduction}

There has been a great deal of interest and activity in the last two years
on the possibility that there may be one or more extra space dimensions in
nature which have sizes of order of a millimeter\cite{arkani}. This has
been driven by the realization that string theories provide a
completely new way to view a multidimensional space time in terms of a
brane-bulk picture, where a brane is lower dimensional space-time manifold
that
contains known matter and forces and the bulk consists of the brane plus
the rest of space dimensions where only gravity is present. The resulting
picture replaces Planck scale by the string scale as the new fundamental
scale beyond the standard model. The relation between the familiar Planck
scale and the string scale $M_*$ is given by the formula\cite{arkani}
\begin{eqnarray}
M^2_{P\ell}~=~M^{2+n}_* R_1R_2R_3 \cdot\cdot\cdot R_n
\end{eqnarray}
For $R_1\simeq R_2=R$ and $R_3\simeq R_4\simeq \cdot\cdot= M^{-1}_*$, this
relation leads to $R\simeq$ millimeter for $M_*\simeq $TeV. The fact that
the familiar inverse square law of gravity allows for the existence of
such sub-millimeter size extra dimensions made these models interesting
for phenomenology\cite{adel}. An added attraction was the fact that a
whole new set of
particles are present at the TeV scale making such theories 
accessible to
collider  tests. Furthermore since there are no high scales in
the theory, there is no hierarchy problem between the weak and the Planck
scale; this provided an alternative approach (resolution ?) to the
familiar
gauge hierarchy problem. Obviously, the picture would become much
more interesting if collider
experiments such as those planned at LHC or Tevatron fail to reveal any
evidence for supersymmetry. 

Even though these models present an attractive alternative to the
standard grand unification scenarios, there are
two arenas where the simplest TeV string scale, large extra dimension
models lead to problems: (i) one has to do with understanding neutrino
masses and (ii) second is in the domain of cosmology and astrophysics.

The reason for the
first is that the smallness of neutrino masses is generally thought to be
understood via the seesaw mechanism\cite{seesaw}, the fundamental
requirement of which is the existence of a scale 
$\geq 10^{11}$ or $10^{12}$ GeV, if the neutrino masses are in the sub eV
range. Clearly this is a much higher scale than $M_*$
of the TeV scale models. A second problem is that if one considers only
the standard model group on the brane, operators such as $LH LH/M_*$ could
be induced by string theory in the low energy effective Lagrangian. For
TeV scale strings this would obviously lead to unacceptable neutrino
masses.

In the domain of cosmology, the problems are
related to the existence of the
Kaluza-Klein (KK) tower of gravitons generally which lead to
overclosure of the Universe unless the highest temperature of the
universe is about an MeV\cite{hanestad}. This can not only cause potential
problems with big bang nucleosynthesis but also with
understanding of the origin of matter, inflation etc. There are also
arguments based on SN1987A observations that require that $M_*\geq 50-100$
TeV\cite{sn}.

The neutrino mass problem was realized early on and a simple solution was
proposed in ref.\cite{dienes}. The suggestion was to postulate
the existence of one or more gauge singlet neutrinos, $\nu_B$, in the
bulk which couple to the lepton doublets in the brane. We will call
this the bulk neutrino. After
electroweak symmetry breaking, this coupling can lead to neutrino Dirac
masses, which are of order $h v_{wk}M_*/M_{P\ell}$. This leads to
$m_{\nu}\simeq 10^{-4}$ eV. The dominant nonrenormalizable terms have to
be forbidden in this model. The simple way to accomplish this is to 
assume the existence of a global B-L symmetry in the theory.
 The only difficulty with this assumption is that string
theories are not supposed to have any global symmetries and one has to
find a way to generate an effective B-L symmetry at low energies without
putting it in at the beginning. 

There is an alternative scenario for neutrino masses\cite{nandi,cald},
where one abandons
the TeV string scale but maintains one large extra dimension and avoids
the problem associated with nonrenormalizable operators. The relation in
Eq. (1) then gets modified to the
form \begin{eqnarray}
M^2_{P\ell} ~=~ M^3_*R
\end{eqnarray}
Since the string scale in these models is in the intermediate range
i.e. $10^{9}$ GeV or so, the cosmological overclosure problem is avoided.
 The active neutrino masses in such
models could arise from seesaw mechanism or from the presence of bulk
neutrinos. The inclusion of the bulk neutrino however brings in new
neutrinos into the theory which can be ultralight (i.e. $R^{-1}$, where
$R$ is the size of the large extra dimension) and can play the role of the
sterile neutrino, which may be required e.g. if the LSND results are
confirmed.

Both the above approaches have the common feature that they introduce a
bulk neutrino into the theory, which is equivalent in the brane to an
infinite tower of sterile neutrinos. All of these neutrino modes mix with
the active neutrinos in the process of mass generation\cite{dienes,nandi}.
For extra dimension size of order $\sim$ mm, the KK modes have masses
typically of order
$nR^{-1} \sim n\times 10^{-3}$ eV or so, where $n=
0,1,2,3,\cdot\cdot\cdot$. The presence of
this dense tower of extra sterile states coupled to the known neutrinos
leads to a variety of new effects in the domain of particle
physics\cite{das} and cosmology\cite{barbieri1,fuller,lukas}, which in
turn impose
constraints on the allowed size of the extra dimensions. In this paper, we
focus on the cosmological constraints that may arise from the
contribution of the neutrino states to big bang nucleosynthesis. 

The constraints from big bang nucleosynthesis on bulk neutrinos were
considered in Ref.\cite{barbieri1,fuller}, where the cases of $\geq$ 2 
extra dimensions were discussed. As it was noted there,
light KK modes of the bulk neutrinos could easily be produced in the early
universe, when temperature is of order an MeV or more.
 Thus, there is
the danger that they could make large contribution to the energy density
of the
universe at the epoch of big bang nucleosynthesis (BBN) and completely
destroy our current, successful understanding of the primordial abundance
of $He^4$, $D$ and $Li^7$\cite{olive}. 
In particular, present abundance data from metal poor stars is well
understood
provided we do not have more than one  extra active neutrino in
the theory in addition to the three known ones i.e. $(\nu_e,
\nu_{\mu}, \nu_{\tau})$. By the same token, 
if there are 
extra species of neutrinos that do not
have conventional weak interactions, their masses and mixings to known
neutrinos must obey severe
constraints\cite{dolgov}.

Our goal in this paper is to revisit this issue in the context of models
with only one large extra dimension. The first reason for undertaking this
analysis is that the class of models with intermediate string scale $\sim
10^{9}$ Gev and one large extra dimension\cite{nandi,cald} have
certain theoretical advantages and they are also free of the cosmological
and
astrophysical problems that seem to plague the TeV scale models. Secondly,
the number of KK modes in this case are much fewer than models with larger
number of large extra dimensions
and therefore, one would expect the constraints to be somewhat less
restrictive. 

We also wish to emphasize that one large extra dimension could also occur
in models with string scale in the 100 TeV range, where one can satisfy 
the Planck-scale-string-scale relation in
Eq. (1), if the compactification is not isotropic, e.g.
for a string scale of 100 TeV, if two extra dimensions have sizes $r \sim$
GeV$^{-1}$ and one has $R\sim $ millimeter, i.e. $M^2_{P\ell}= M^5_* R
r^2 $.

We organize this paper as follows: in sec II, we discuss the
constraints of big bang nucleosynthesis on the size of the extra dimension
in the presence of the bulk neutrino and the mixing of the bulk neutrino
to the active one using simple out of equilibrium condition. In section
III,
we use Boltzman equation to study the generation of the bulk neutrinos
from active neutrino interactions in the early universe and find the
constraint of BBN on the same parameters as in sec. II. The numerical
calculations leading to our final results are given in sec.IV. In 
appendix A,
we explain the details of the out of equilibrium condition for
the KK modes of the neutrino.

 \bigskip

\section{Constraints from out of equilibrium condition}

The class of models, we will be interested in, are assumed to have one
large extra dimension with a single bulk neutrino $\nu_B$, which means
that masses of the KK modes of $\nu_B$ are integer multiples of the basic
scale $\mu \equiv R^{-1}\sim 10^{-3}$ eV. This is one of the parameters
that
we expect the BBN discussion to constrain. The second parameter is the
mixing of the KK modes with the active brane neutrinos, e.g. $\nu_e$. It
is true in both classes of models i.e. both
TeV and intermediate scale type,\cite{dienes,cald} that the typical
mixing parameter of the active neutrino to the nth KK mode
scales like $\theta_{en}\simeq \frac{\theta}{n}$, in the range of
interest for phenomenology. The parameter $\theta$
depends on the size of the extra dimension and other parameters of the 
theory such as the weak scale etc. For instance, in TeV scale
models\cite{dienes}, one has $\theta\sim
\frac{\sqrt{2}hv_{wk}M_*R}{M_{P\ell}}$, whereas in the local B-L
models, the relation is $\theta\simeq
\frac{hv_{wk}v_RR}{M_{P\ell}\mu}$, where we have chosen $M_*\simeq
v_R$. In general, therefore,
BBN discussion will give a correlated constraint between $\mu$ and
$\theta$. Obviously, for $\theta = 0$, there is no BBN constraint on
$\mu$ or the size of the extra dimension. 

We will also assume that the universe starts its ``big bang journey''
somewhere around a GeV or so and when it starts, the universe is
essentially swept clean of the sterile neutrino modes. This can happen in 
inflation models with a low reheat temperature. We choose such a low
reheat temperature essentially for reasons that in models with large extra
dimensions higher temperatures would lead to closure due to production of
graviton KK modes.

To see the origin of constraints, let us note that at high temperatures
(i.e. $T\gg $ MeV's), there are two ways the KK modes of the sterile bulk
neutrino can be
created: (i) first, neutrino scattering and annihilations and (ii) the
oscillation of the active neutrinos
into the sterile KK modes. It is important to stress that in building up
the oscillation, the scattering process is important, since otherwise
there will be back-and-forth oscillation and no build-up of the sterile
modes.

 Since there is an infinite KK tower of these neutrino
modes, the higher the temperature the larger the number of modes that can
get created. Once these modes are created, they may decay or annihilate to
produce the lighter particles (lighter neutrino modes or KK modes of
the graviton etc). In general, it is reasonable to expect that
this process of decay or annihilation will not be efficient
enough\cite{fuller} to eliminate all the KK modes. As a result, many of
them will stay around at the BBN temperature and contribute to the energy
density. The present understanding of the big bang
nucleosynthesis\cite{olive} relies
on the assumption that the total energy density at the BBN era is
$\rho_{BBN}~=~\frac{\pi^2}{30}g^*
T^4$ with $g^*=10.75$ coming from the contribution of photon, $e^+e^-$,
and the three species of neutrinos. The uncertainties in our knowledge of
the $He^4$, $D_2$ and $Li^7$ content of the universe allow that one could
have $g^*\sim 12.5$ (or one extra species of neutrino). We will require
that any additional contribution to $\rho_{BBN}$ coming from the bulk
neutrinos generated at higher temperature be less than the
contribution to $\rho_{BBN}$ equivalent to one extra species of neutrino.

The first step to ensure this would be to enforce the condition
that the production rate for any singer bulk neutrino mode is less than
the Hubble 
expansion rate of the universe between 1 GeV to MeV. This will prevent any bulk 
neutrino mode from equilibrium. The equilibration of any neutrino mode is
unacceptable because
it can contribute more than allowed energy density to the
universe. Clearly this condition cannot be a completely reliable method of
constraining the parameter space when there are such a large number of
final
states present since a small oscillation into each mode can ruin BBN
results. However, we use this as a ``warm up'' to our final (hopefully
more refined treatment) and as a basic guideline for 
more precise and stronger condition. We present the details of this
discussion and its
results in this section fully realizing that this is not going to be the
``final story''. Our next step to obtain constraints is to use the
Boltzman equations for the time evolution of the density of the KK modes
and obtain their contributions to the energy density at the BBN era and
demand that this energy density is less than that of one extra species of
neutrino. We will discuss this in the next section. We find it intriguing
that both these methods lead to bounds which are very close to each other.

We begin with some pedagogical comments by considering one sterile mode
which mixes with one of the active neutrinos and by ignoring the
matter effects. We present this discussion as a way to appreciate the
importance of the matter effects. In this simple case,it turns out that it
is the oscillations rather than direct production in scattering that
give the strongest bound\cite{fargion}.

To see that direct production is not so important, note that the
the production rate for the nth KK mode is roughly given by
$G^2_FT^3\mu^2\theta^2$. Note that dependence on the mode number cancels
out. Setting this $\leq $ the Hubble expansion rate , we get,
for the decoupling temperature $T^*\simeq
g^{*1/2}/(G^{2}_F\mu^2\theta^2M_{P\ell})\simeq
10^{11}\frac{GeV}{\theta^2}$. Thus below this temperature all production
processes for arbitrary modes are far out of equilibrium. Thus we only
have to consider $\nu_B$ production through oscillations.

The transition rate to one sterile (or KK) mode from oscillation in the
absence of matter effect can be estimated as follows: the
oscillation is a quantum mechanical phenomenon that gets interrupted as
soon as a collision takes place. The amount of time which allows a build
up of the oscillation to the sterile state is therefore given by the
time between the collisions $\tau$ i.e. inverse of the collision rate,
$W_{wk}
\simeq G^2_FT^5$. The probability for transition to a sterile 
mode in time $\tau \simeq W^{-1}_{wk}$ is given by 
\begin{eqnarray}
P\simeq sin^22\theta_{en}~ sin^2 \left(\frac{\tau_{wk}}{\tau_{osc}}\right)
\end{eqnarray}
where oscillation time $\tau_{osc}\simeq \frac{T}{\Delta m^2}\simeq
\frac{T}{m^2_{KK}}$.
If time between collisions of the $\nu_e$, denoted
by $\tau_{wk}$, is much less than the oscillation time, this expression
is simplified and we get  the
production rate for the sterile KK modes
\begin{eqnarray}
W_{KK} \simeq P W_{wk}\simeq
sin^22\theta_{en}\left(\frac{\tau_{wk}}{\tau^2_{osc}}\right)
\end{eqnarray} 
 This condition
i.e. $\tau_{wk}\equiv G^{-2}_FT^{-5} \leq E/\Delta m^2$ is satisfied
around 2-3 MeV for $m \leq 10^{-3}$ eV. Below this temperature we must
approximate the oscillation factor by 1/2. Staying above 3 MeV and 
ignoring the matter effects, we can implement the
BBN constraint by requiring that the production rate for a KK mode be less
than the Hubble expansion rate at a given epoch:
\begin{eqnarray}
W_{KK} \leq {g^*}^{1/2}\frac{T^2}{M_{P\ell}}
\end{eqnarray}
If a given KK mode is much lighter than an MeV, we get the bound in
literature\cite{fargion} that
\begin{eqnarray}
m^4_{KK}\theta^2_{en} \leq 10^{-19}~~eV^4
\end{eqnarray}

Taking matter effects into account essentially amounts to replacing the
vacuum mixing angle by
the matter mixing angle $\theta^m_{en}$ given by
\begin{eqnarray}
sin^22\theta^m_{en} = \frac{sin^22\theta_{en}}{1-2zcos2\theta_{en}+ z^2}
\end{eqnarray}
where $z=-6.3 T\frac{\sqrt{2}G_Fn_{\gamma}55 T^2}{m^2_Wm^2_{KK}}\simeq
-\frac{2.155 \times 10^{11}(T/GeV)^6}{k^2(\mu/eV)^2}\equiv
-\frac{\alpha(T)}{k^2}$
for the $k$-th KK mode\cite{notzold}. We have assumed that there is 
no lepton asymmetry. Clearly
$\alpha(T)$ is a function of
temperature. Note that $z$ is always negative. This means that
there is no resonance type behavior for the mixing
angle\footnote{ Note that if there was a
significant initial lepton asymmetry, the situation would have been very
different and whether there is a resonance would depend on whether the
initial particle is a neutrino or an antineutrino.}, which makes the
calculation more reliable.

 The out of equilibrium
condition in the presence of matter effect is given by:
\begin{eqnarray}
\frac{sin^22\theta}{k^2(1-z)^2}sin^2\frac{\mu^2k^2}{24T^6G^2_F}
\leq 2.5 \sqrt{g^*}T^{-3}10^{-10}
\end{eqnarray}
where all mass parameters are in units of GeV; $\mu= R^{-1}$ is the KK
scale, discussed before. This condition must be satisfied for all $k$ and
within our
temperature window of one MeV to one GeV. We include the detail analysis of this 
part in Appendix A. For $\theta, \mu, T$ fixed,
the equation has to be satisfied when the expression in the left hand side
of the above expression is a maximum. This occurs when $k =
\sqrt{\alpha(T)}$
for $\alpha (T)\geq 1$. For $\alpha (T) < 1$, the maximum occurs when
$k=1$. In
the first case, we get the limit,
\begin{eqnarray}
\mu^2 sin^22\theta \leq 2.15 (\frac{T}{GeV})^3 \sqrt{g^*} 10^2 eV^2
\end{eqnarray}
The best bound arises when $T$ is minimum consistent with the condition
$\alpha (T) \geq 1$ or equivalently
\begin{eqnarray}
(\frac{T}{GeV})^6 \geq \frac{(\mu/eV)^2}{2.15\times 10^{11}}
\end{eqnarray}
In the second case, we get the limit,
\begin{eqnarray}
sin^22\theta \leq 2.5 (\frac{T}{GeV})^{-3} \sqrt{g^*} (1+\alpha)^2 10^{-10}
\end{eqnarray}
The best bound arises when $T$ is maximum consistent with the condition
$\alpha (T) < 1$ or equivalently
\begin{eqnarray}
(\frac{T}{GeV})^6 < \frac{(\mu/eV)^2}{2.15\times 10^{11}}.
\end{eqnarray}
 The temperature should also be within our chosen
window of one MeV to GeV. This gives rise to three cases. To see the
various cases, let us define the temperature $T_1 =
\left(\frac{(\mu/eV)^2}{2.15\times 10^{11}}\right)^{1/6} GeV$ . The various
cases then correspond to (i) $T_1 < $ MeV; (ii) GeV $< T_1$; 
(iii) MeV $< T_1 \leq $ GeV. The limits for various cases are found to be as
follows: \\
\\
(i) In this case $\mu^2$ lies in the range $\mu^2 < 2.15\times 10^{-7}
eV^2$ \\ and the constraint on $sin^22\theta$ is given by
\begin{eqnarray}
\mu^2~ sin^22\theta < 7.07 \times 10^{-7} eV^2.
\end{eqnarray}
As $sin^22\theta$ is less than 1, this imposes no further
constraint on our parameter space.\\
(ii)For this case, $\mu^2 \geq 2.15 \times 10^{11}$ eV$^2$ Which is far beyond our range 
of interest. \\
(iii) $\mu^2$ is in the range $2.15\times 10^{-7} eV^2 <
\mu^2 < 2.15 \times 10^{11}$ eV$^2 $ \\
and the bound is
\begin{eqnarray}
\mu~ sin^22\theta < 5.804 \times 10^{-4} eV
\label{equilibrium}
\end{eqnarray}

The bound of the allowed parameter space due to the out of equilibrium condition is 
adequately described by eq. (\ref{equilibrium}) within the range of interest.   
We plot the result in Fig. 1. It corresponds to the small dashed line, 
the right side of which is forbidden.

 \section{Boltzman Equation and BBN constraints}
In this section, we employ the Boltzman equations to get the constraints
on $\mu$ and $\theta$. Our procedure is to calculate the distribution
function for the sterile KK modes produced in the matter oscillation of
$\nu_e$, including any possible depletion of their density due to decays
all the way down to the BBN temperature. We then calculate their
cumulative contribution to the energy density $\rho$ at the BBN epoch and
demand that this be less than the corresponding contribution of one extra
species of neutrino. This is the procedure followed in
\cite{Dodelson,fuller}.

To estimate this new contribution to energy density $\rho_{BBN}$, we have
to calculate the number of KK states produced at a given temperature. Let
us denote by $f_k = f_k(p,t)$, the distribution of the $k$th mode of the
neutrino at the epoch $t$. The KK modes are not in equilibrium at any
epoch.
The time evolution of the nonequilibrium density is governed by the  
Boltzmann equation given below \cite{fuller} \cite{Dodelson}.
\begin{equation}
(\frac{\partial}{\partial t}-H{p}\frac{\partial}{\partial
p})f_k=\Gamma(\nu_{\alpha}\longrightarrow\nu_k)f_{\nu_\alpha}-
{m_k\over E_k} {1\over{\tau_k}}f_k+\sum_{l>k}C_{k,l}[f_l],
\label{bolteq}
\end{equation}
where we have neglected the contribution of the pair annihilation
of the KK modes in the right hand side, since it is a very small effect.
 In Eq. (\ref{bolteq}), $t$ (time), $p$ (momentum) are
the two independent variables with $k$, $\mu$ and $sin^22\theta$ fixed. ${{E_k}}(p)
 =\sqrt{p^2+{m_k}^2}$. $H$ is the instantaneous Hubble expansion rate and
is clearly  a function of time $H(t)$.
$\Gamma(\nu_{\alpha}\longrightarrow\nu_k)=\Gamma_k$, the
production rate of  bulk neutrino, 
 is given by $(\Gamma /2) \langle
P(\nu_{\alpha}\longrightarrow\nu_k)  \rangle$ where $(\Gamma /2)=2 {G_F}^2
T^5$ is half the interaction rate of the active neutrino in the thermal
bath. Taking matter effects into account, the probability $P$ is given by
 \begin{equation}
 \label{prob}
\langle P (\nu_\alpha\rightarrow \nu_k) \rangle \simeq
{\frac{1}{2}} \frac{\sin^2
2\theta_k}{1 - 2z\cos2\theta_k+z^2}.
\end{equation}
Note that we have used the averaged probability $P$ to get rid of the momentum dependence 
of the production rate. We will only use this explicit form in our numerical calculation. 
The averaging is not necessary for the general solution we are going to find later 
in this section. In the above equation, $f_{\nu_\alpha}$ is thermal distribution of active 
neutrino $\nu_\alpha$. $\tau_k$ is the lifetime of the $k$-th mode in its rest frame, 
and $\frac{1}{\tau_k}=$ total decay width. For small $k$, the dominant
contributions come from the  partial width of $\nu_k \rightarrow 3 \nu$
decay which is given by 
$sin^2 \theta_k G^2_f m^5_k/192 \pi^3 = sin^2 2\theta_k G^2_f \mu^5
k^5/(4 \times 192 \pi^3)$\cite{fuller}.
For big $k$, it is from $\sum_{k'=1}^{k-1} \nu_k \rightarrow
\nu_{k'}h_{k-k'}\sim m_k^3(k-1)/12\pi M_{pl}^2
\sim k^4\mu^3/12\pi M_{pl}^2$. We included both contributions in our
total decay rate. To make the solution more general, we now let all of the functions in 
Eq. (\ref{bolteq}) except $H(t)$ depend on both $p$ and $t$. 

The general solution of eq.   (\ref{bolteq}) without knowing the form of
functions introduced in
 the equation and $C_{k,l}=0$ is found to be
 \begin{equation}
 \label{bolzsol}
 f_k (p,t) =\int_{t_i}^{t}  \Gamma_k (p'(x,t),x) f_{\nu_\alpha} (p'(x,t),x) 
e^{-m_k\alpha(x,t)}dx
 \end{equation}
 provided  
 \begin{equation}
 p' (x,t) =p  e^{\int _{x}^{t}H(x') dx'}		
 \end{equation}
Note from Eq. (18) that $p'(t,t) = p$.
\begin{equation}
\alpha(x,t) =\int _{x}^{t} \frac{1}{\tau_k(p'(x',t),x')\sqrt{p'^2(x',t)+{m_k}^2}} dx'
\end{equation} 
 The above solution can be checked easily by the observation that
  \begin{equation}
 (\frac{\partial}{\partial t}-H{p}\frac{\partial}{\partial p}){p'}(x,t) =0
  \end{equation}
and 
  \begin{equation}
 (\frac{\partial}{\partial t}-H{p}\frac{\partial}{\partial p})F({p'}(x,t)) =0
  \end{equation}
for any function $F({p'}p (x,t))$ with no explicit dependence on $t$ and
$p$. 
The time derivative on the integration upper limit gives the first term
on the right 
hand side of eq. (\ref{bolteq}). While acting on the upper limit of $\alpha(x,t)$, it 
gives the second term on the right hand side of eq. (\ref{bolteq}).
The total energy density of bulk neutrino is then given by 
\begin{equation}
\label{density}
\rho(t)=\int_{0}^{\infty}\int_{1}^{\infty} \frac{E_k p^2}{2 \pi^2} f_k
(p,t) dk dp
 \end{equation}
in the continuous $k$ approximation. In the next subsection, we
discuss the numerical results that follow from
the above discussion.

\section{Numerical results}
 
 In this section, we calculate the total energy density of the universe at
$T=1MeV$ from (\ref {density}).  
 We use temperature (T) instead of time in the integration and proceed 
 in two different ways. 

In the first method, we make some approximations to simplify the integral
in Eq. (17)
 and calculate  analytically as far as possible and then numerically
estimate the final integrals that follow from it.

In the second method, we evaluate the entire integral in Eq. (17) 
numerically . The two results agree approximately with each other.
  
In the calculation, we use $H(T)=1.66\sqrt{g_*}\frac{T^2}{M_{pl}}$ and 
$t=0.301 g_*^{-\frac{1}{2}}\frac{M_{pl}}{T^2}$ as given by Standard cosmology.  
After introducing dimensionless variables $p'=\frac{p}{k \mu}$,  
$x=\frac{T}{T_f}$ and $\alpha = k \frac{\mu}{T_f} $,and using the
notation $p$ instead of $p'$ 
in the final form, Eq. (\ref {density}) reduces to
\begin{equation}
\label{denb}
\rho(T_f)={\cal A}\int_{1}^{\frac{T_i}{T_f}}\int_{\epsilon}^{\infty} 
\int_{0}^{\infty}
\frac{\alpha^2 x^2 p^2 \sqrt{1+p^2}(1+\frac{x}{4}\frac{d\ell n g_*(x)}{dx})}
{\sqrt{g_*}(1+z^2-2z\sqrt{1-\frac{sin^22\theta \mu^2}{\alpha^2T^2_f}})}
e^D {\cal
F} dp d\alpha dx,
\end{equation}
 where
\begin{equation}
 {\cal A}=\frac{0.301}{\pi^2}\mu sin^22\theta G_F^2 M_{pl} T_f^6;
\end{equation}
\begin{equation}
  D~=-\frac{0.301 M_{pl}}{\sqrt{10.75}}[\frac{\mu^2 sin^2 2\theta
G_F^2T_f}{4\times 192 \pi^2}\alpha^3
  +\frac{T_f^2}{12\pi M_{pl}^2
\mu}\alpha^4]K(x,p)
\end{equation}
 where $K(x,p)~=~
(\sqrt{1+p^2}-\frac{c^{-\frac{1}{4}}}{x}
\sqrt{p^2+\frac{c^{-\frac{1}{2}}}{x^2}}-p^2 \ell n 
\frac{1+\sqrt{1+p^2}}{\frac
{c^{-\frac{1}{4}}}{x}+\sqrt{p^2+\frac{c^{-\frac{1}{2}}}{x^2}}})$.
\begin{equation}
  {\cal
F}=\frac{1}{e^{\alpha p c^{\frac{1}{4}}}+1}.
\end{equation}
and  $c=\frac{g_*(x)}{g_*(T_f)}$, $\epsilon=\mu/T_f$.
In order to simplify the calculation, we also make the following
approximations. \\

$\bullet$ We assume that the effective degree of freedom $g_*$
is not affected by the density of the bulk neutrino during the production
of bulk neutrino. 
$g_*$ is given by the standard model of cosmology and approximated by  a step
function as follows: 
$g_*=61.75$ from  $T=1 GeV$ to $200$ MeV, $g_*=17.5$ from  $T=  200$ MeV
to $100$ MeV, 
and $g_*=10.75$ from $T=100$ MeV to $1$ MeV.  
Here we actually
approximate the degree of freedom as 
a step function of temperature.\\

$\bullet$ We use $\delta m^2 \equiv m_k^2 - m_\nu ^2 \approx k^2\mu ^2$ \\

Our $\delta m^2$ are always positive and so lepton number generated from
neutrino oscillation with nonzero 
initial lepton number is small \cite{FootLep}.  We can ignore the
contribution of lepton number
to the $\nu _e$ potential in matter. This gives
\begin{equation}
 z=-0.215589\times
10^{-18}\frac{x^6}{\alpha^2}
\end{equation}
$c$ is a constant of O(1) in the step
function approximation of $g_*$.

 In order to use the analytical method we make some further
approximations. Note that the integration in Eq. (23) is
 suppressed by the two exponential functions: one from the decay term ($D$
term) $e^D$ and a second from the ${\cal F}$ term. We can therefore cut
off the integration when
 either one of them gets smaller than $e^{-100}$.  We want to extract
some of the regions 
 of $\alpha$ and $p$ space in which the ${\cal F}$ term  
 decays much faster than the $D$ term within the parameter space of 
interest and so the latter 
can be ignored. However, it is easier to do it another way. As the  
contribution from the region where 
${\cal F} < e^{-100}$ can be neglected, we have to examine only region
where $p \alpha <100$ to make sure 
$D$ term can be approximated to 1.
This is same as finding the region where $p \alpha <100$ and both $\frac{4
k^4}{\mu} \times 10^{-19}$ and 
$64 k^3\mu^2 \sin^22\theta\times 10^{-19}$ less than 0.1. It is easy to
see that the only region that may not 
satisfy the above condition is 
\begin{equation}
\alpha > min[(\frac{\mu 10^{18}}{4})^{\frac{1}{4}},(\frac{10^{18}}{64
\mu^2 \sin^22\theta})^{\frac{1}{3}}] \equiv k_{min}
\end{equation} 
and 
\begin{equation}
p<\frac{1}{k_{min}} < 1
\end{equation}

Now we can calculate the total energy density by splitting the
integration into three parts : \\
(i)  $p > 1$ \\
(ii) $p <1 $ and $\alpha < k_{min}$ \\
(iii) $p <1$ and $\alpha > k_{min}$ \\

We will include the decay term only for (iii). We also simplify our
analysis by substituting 
 ${\cal F}= \frac{1}{2}e^{-\alpha p }$ with a
possible error of factor 2.
 For simplicity, we will now treat $g_*$ as a constant and so
$c=1$. Setting $cos2\theta=1$ affects  only the 
term due to the matter effect. This is a very good approximation as long
as $z$ is negative. For cases (i) and (ii), Eq. (\ref {denb}) now becomes

\begin{equation}
\label{denc}
\rho=\frac{\cal
A}{2\sqrt{g_*}}\int_{1}^{10^3}\int_{\epsilon}^{\infty}\int_{0}^{\infty}
\frac{\alpha^2 x^2 p^2 \sqrt{1+p^2}}
{(1-z)^2} e^{-\alpha p } dp d\alpha dx
\end{equation}

The $x$ integration can now be done analytically. we approximate
$\sqrt{1+p^2}$  with $p$ 
for case (i) where $p > 1$ and with $1$ for case (ii) where $p < 1$. With
this simplification, 
the $p$ integration can also be done analytically. For $\alpha$, we can always 
divide it into three parts: $ \alpha > 10$ , $ 0.01<\alpha < 10 $ and
$\alpha < 0.01$ since 
$k_{min} > 10^3$.   we will leave the factor $\frac{\cal
A}{\sqrt{g_*}}$ to the end of the discussion.Now we only look at the
integration. 
The results was summarized below

\bigskip

\noindent (i) $p > 1$ .\\
For $\alpha>10$, we have 

\begin{equation}
\frac{10^9}{6} \int_{10}^{\infty} e^{-k}k dk  = 83233
\end{equation}
For $0.01 < \alpha < 10$, we have

\begin{equation}
10^9\int_{0.01}^{10}
e^{-k}(\frac{1}{2k^2}+\frac{1}{2k}+\frac{1}{4}+\frac{k}{12})
 (\frac{1}{1+\frac{0.216}{k^2}}+\frac{\tan
^{-1}\frac{0.465}{k}}{\frac{0.465}{k}}) dk  
= 7.85 \times 10^9
\end{equation}
For $\alpha < 0.01$

\begin{eqnarray}
\frac{1}{2}\int_{\epsilon}^{0.01}
\frac{10^9}{0.216}+\frac{\pi}{2ka}-\frac{1}{k^2+a^2}
-\frac{\tan ^{-1}\frac{a}{k}}{ka} dk  \\
= \frac{10^7}{0.432} + \frac{\pi}{4a}\ln
\frac{0.01}{\epsilon}-\frac{\frac{\pi}{2}-
\tan^{-1}\frac{\epsilon}{a}}{2a}- \\
\frac{1}{2a}\int_{\epsilon}^{0.01}\frac{\tan^{-1}\frac{a}{k}}{k} dk
\end{eqnarray}

\bigskip

\noindent (ii) $p < 1$ and  $\alpha < k_{min}$.\\
For $\alpha>10$, we have 

\begin{equation}
\frac{10^9}{3} \int_{10}^{k_{min}}\frac{1}{k} dk  = \frac{10^9}{3}\ln
\frac{k_{min}}{10}
\end{equation}
For $0.01 < \alpha < 10$, we have

\begin{equation}
10^9\int_{0.01}^{10} (\frac{1}{6k}-e^{-k}(\frac{1}{6k}+\frac{1}{6}+\frac{k}{12})
(\frac{1}{1+\frac{0.216}{k^2}}+\frac{\tan
^{-1}\frac{0.465}{k}}{\frac{0.465}{k}}) dk  
= 4.52 \times 10^8
\end{equation}
For $\alpha < 0.01$ we have 

\begin{equation}
\frac{1}{36}\int_{\epsilon}^{0.01} \frac{10^9 k^4}{0.216}+\frac{\pi k^3}{2a}-\frac{k^4}{k^2+a^2}
-\frac{k^3\tan ^{-1}\frac{a}{k}}{a} dk  
\end{equation}
This is a very  small number which can be neglected.

\bigskip

\noindent(iii) In this region, the decay term have to be
included. However, we can remove the $p$ dependent 
by setting $p=1$ in the function $K(x,p)$, defined above. The matter
effect can also be neglected as $k_{min} > 500$. After the $p$
integration, with the fact that 
$e^{-k_{min}} \leq 1$ we have the upper bound
\begin{equation}
\int_{1}^{10^3}\int_{1}^{\infty}
 x^2 e^{-(\alpha k^4+\beta k^3)K(x, 1)} dk dx
\end{equation}
In the above equation, $\alpha$
and $\beta$ are 
coefficients dependent on $k_{min}$. One of them is $0.1$ and the other
$\leq 0.1$ by the definition of 
$k_{min}$. We also use $k=k_{min}$ in the result of the $p$
integration. For the upper bound, we can 
just take either the $\alpha$ or the $\beta$ term whichever have the
value of $0.1$. this gives 

\begin{equation}
\int_{1}^{10^3}
 x^2 \frac{\Gamma(\frac{1}{4},0.1K(x,1))}{4(0.1K(x,1))^{\frac{1}{4}}} dx
\end{equation}
or 
\begin{equation}
\int_{1}^{10^3}
 x^2 \frac{\Gamma(\frac{1}{3},0.1K(x, 1))}{3(0.1K(x, 1))^{\frac{1}{3}}} dx
\end{equation}
both give numerically order of $10^8$. Our result from other parts give
about $10^{10}$.
This part contributes only a few percent of the total. We can use
the upper bound $4.6\times 10^8$.  

The total energy density of bulk neutrinos is the sum of all above ,
which is found to be

 \begin{equation}
\rho=\frac{\cal
A}{\sqrt{g_*}}(8.785 \times 10^9+ \frac{\pi}{4a}\ln \frac{0.01}{\epsilon}-
\frac{\frac{\pi}{2}-
\tan^{-1}\frac{\epsilon}{a}}{2a}-\frac{1}{2a}\int_{\epsilon}^{0.01} 
\frac{\tan^{-1}\frac{a}{k}}{k} dk+
\frac{10^9}{3}\ln \frac{k_{min}}{10}) \\
\end{equation}
Compared to the $\nu_e$ equilibrium energy density
at $T=1MeV $ which is $2.824\times 10^{-13} (GeV)^4 $, the constraint
that effective neutrino degree of freedom 
should be less than one, gives the constraint on the parameter space
\begin{eqnarray}
\left(\frac{\mu}{eV}\right)^{0.936}~ \sin^22\theta  \leq 2.5 \times
10^{-4} 
\end{eqnarray}

\begin{figure}[h!]
\begin{center}
\epsfxsize15cm\epsffile{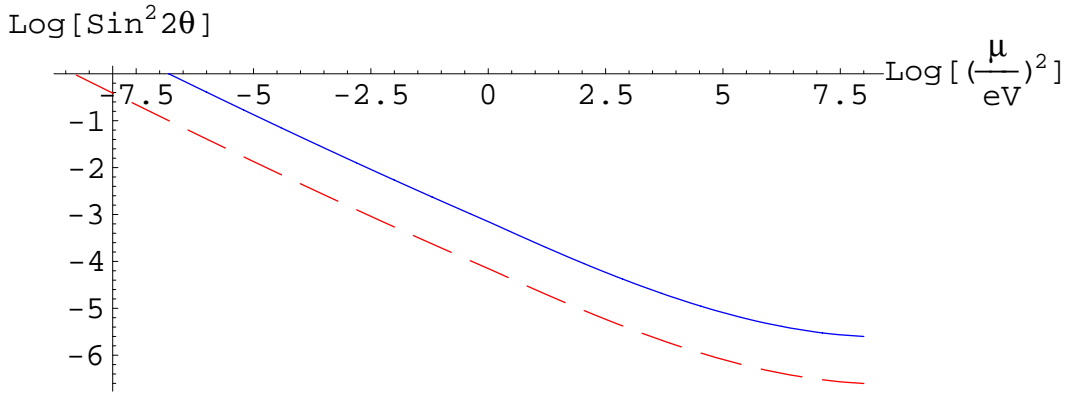}
\caption{ The solid and the long dash line are the numerical results
which represent the effective 
degree of freedom equal to 1 and 0.1 respectively. The small dash line
separate the equilibrium and 
out of equilibrium  region. The parameter space below the lines are the
allowed region.
\label{fig:cstr1}}
\end{center}
\end{figure}

We also numerically integrate e.q (\ref {density}) without making any
simplification except the step function $g_*$ 
and the positive $\delta m^2$.  Both results from the Boltzman equation 
and the
out of equilibrium condition are shown 
in Fig.\ref {fig:cstr1}.The extra effective degree of freedom equal to 1
is shown in Fig. 1 as a solid line. The numerical fit 
is obtained to be
\begin{eqnarray}
\left(\frac{\mu}{eV}\right)^{0.92}~ sin^22\theta  \leq 7.06 \times
10^{-4} 
\end{eqnarray}
for $\mu \leq 1 eV$. 

\bigskip

\bigskip

We note that the bounds derived in both ways i.e. out of equilibrium
condition and Boltzman equation are very similar.

In Fig. 2, we give the contributions of individual modes to the total
energy density as a function of the mode number for $sin^22\theta =
10^{-4}$ and for a range of values for $R^{-1}\equiv \mu $ from $10^{-6}$
eV to $10^{4}$ eV in increasing steps of $10^2$ eV. The total contribution 
is the integral over each line for a given $\mu$.

\bigskip
\bigskip

 \begin{figure}[h!]
\begin{center}
\epsfxsize15cm\epsffile{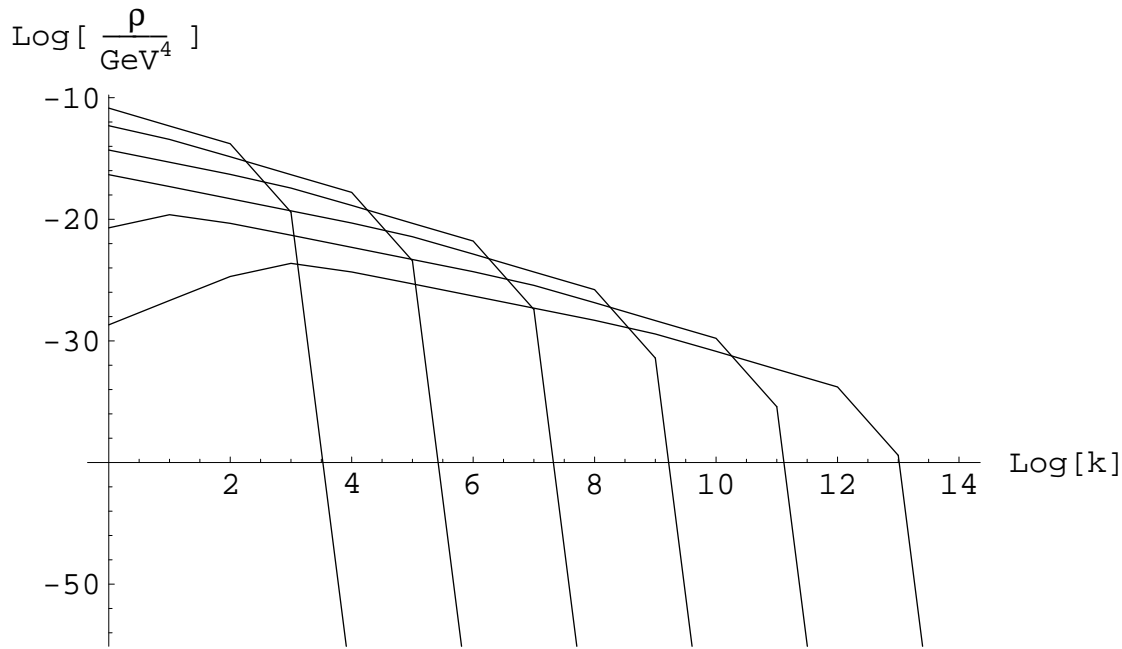}
\caption{ The various lines denote the contribution of the individual
modes to the total energy density as a function of the KK mode number
for various values of inverse size (in eV) of the extra dimension. The
right-most curve corresponds to $\mu =10^{-6}$ eV and for each successive
curve, $\mu$ goes up by a factor of 100.
 \label{fig:cstr2}}
\end{center}
\end{figure} 

\bigskip

\section{Conclusion}
In this paper, we have studied the constraints of big bang nucleosynthesis
on the models with one large extra dimensions and a single bulk
neutrino. We find that these bounds allow a range
of radius of extra dimension and mixing of bulk to active neutrinos that 
is of interest for studying solar neutrino oscillations\cite{cald}. Our
results complement the work of Abazajian, Fuller and Patel\cite{fuller}
who have derived bounds for the case four, five and six extra dimensions.
We find it intriguing that the bounds obtained from the naive and crude
``out of
equilibrium'' conditions (given in Eq. (14)) are so similar to the ones
obtained from more
detailed considerations based on the density matrix equations.

\newpage
\section*{Appendix A}
In this appendix, we give details of the out of equilibrium condition used
in section II to get the bounds on $\mu^2$ and mixing $\theta$. Let us
first recall that
the out of equilibrium condition is given by:
\begin{eqnarray}
\frac{sin^22\theta}{k^2(1+z)^2}sin^2\frac{\mu^2k^2}{24T^6G^2_F}
\leq 2.5 \sqrt{g^*}T^{-3}10^{-10}
\end{eqnarray}
where all masses are in GeV units and
\begin{eqnarray}
 z= \frac{2.15\times10^{-7}T^6}{k^2 \mu^2}\equiv \frac{\alpha}{k^2}
\end{eqnarray}
Putting in the numerical value of $M_{pl}$ and $G_F$ we get
\begin{eqnarray}
\frac{sin^22\theta}{(k+\frac{\alpha}{k})^2}sin^2(\frac{66.07}{\alpha}k^2)
\leq 2.5 \sqrt{g^*}T^{-3}10^{-10}
\end{eqnarray}
This condition must be satisfied for all $k$ and within our
temperature window of an MeV to one GeV. For $\theta, \mu, T$ fixed,
if the inequality is satisfied when the expression in the left hand
side is a maximum, clearly it is then always satisfied. The $k$ dependent
part is given by
\begin{eqnarray}
 \frac{sin^2(\frac{66.07}{\alpha}k^2)}{(k+\frac{\alpha}{k})^2}= 
\frac{sin^2( 66.07y)}{\alpha y(1+\frac{1}{y})^2} \\
 y\equiv \frac{k^2}{\alpha}
\end{eqnarray}
Here $\alpha$ is independent on $k$.
In this expression, $ sin^2( 66.07y)$ oscillates so fast  
that in calculating the maximum, we can  set $
sin^2( 66.07y)=1$ and finding the maximum of
$\frac{1}{y(1+\frac{1}{y})^2}$. The maximum is found at $y=1$. $y=1$ implies
$k=\sqrt{\alpha}$. Because $k$ must be positive integer,
we should choose $k$ equal to the closest integer of $\sqrt{\alpha}$ if
$\alpha \geq 1$. For $\alpha < 1$, since $y \geq 1$c and it is easy to see that the maximum 
occurs at the minimum of y, $k=1$ is the maximum point. It is worth noting that,
the condition have to be satisfied within a range of temperature (and so
a continuous range of $\alpha$).Although at some temperature 
with a well chosen $\mu$, the constraint can be weaker than  those we
use, small range of temperature of MeV will cover the period 
of $ sin^2( 66.07y)$. So the constraint will be that obtained just by
setting $ sin^2( 66.07y)=1$.

Now we discuss the two different cases separately. \\
(i) For $\alpha \geq 1\Rightarrow T\geq
(\frac{\mu^2}{2.15\times10^{-7}})^{\frac{1}{6}} \equiv T_1$.we have constraint 
\begin{eqnarray}
\mu^2~ sin^22\theta \leq 2.15 (\frac{T}{GeV})^3 \sqrt{g^*} 10^2 eV^2
 \label{alpha1}
\end{eqnarray}
(ii) For $\alpha<1 \Rightarrow T<
(\frac{\mu^2}{2.15\times10^{-7}})^\frac{1}{6} $. We have constraint
\begin{eqnarray}
sin^22\theta \leq 2.5 (1+\alpha)^2 (\frac{T}{GeV})^{-3} \sqrt{g^*}
10^{-10}\simeq 2.5 (\frac{T}{GeV})^{-3} \sqrt{g^*} 10^{-10}
\label{alpha2}
\end{eqnarray}
In case 2, we have approximate $\alpha \ll 1$. Without making this
approximation, the result will give approximately a factor of 2 to
$sin^22\theta$. Under
this approximation, three different situation appear. they are
(i) $T_1 < MeV \Rightarrow \mu^2 < 2.15\times 10^{-7} eV^2$ ,(ii) $MeV <
T_1 < GeV \Rightarrow 2.15\times 10^{-7} eV^2
< \mu^2 < 2.15 \times 10^{11}$ eV$^2$,(iii) $T_1 \geq GeV \Rightarrow
\mu^2
\geq 2.15 \times 10^{11}$ eV$^2$. \\
(i) $T_1 < MeV$

We have to use eq.(\ref{alpha1}) with $T=1 MeV$ and $g=10.75$. This will
give 
\begin{eqnarray}
\mu^2 sin^22\theta < 7.07 \times 10^{-7} eV^2
 \end{eqnarray}
 (ii)$MeV < T_1 < GeV$
 
 We have to use eq.(\ref{alpha1}) for T $\in (T_1, GeV)$ and use
eq.(\ref{alpha2}) for T $\in (MeV, T_1)$ with $T=T_1$ for both cases. 
 This gives respectively
\begin{eqnarray}
\mu~ sin^22\theta < 2.322 \times 10^{-3} eV
\end{eqnarray}
and
\begin{eqnarray}
\mu~ sin^22\theta < 5.804 \times 10^{-4} eV 
\end{eqnarray}
It is easy to see that even without the approximation that $\alpha \leq 1$,
the second equation can go up as high as $1.78 \times 10^{-3} eV$ when  $T_1 >  1.2 MeV$.
It is still more stringent than the first.We use the second equation
with $\sqrt{g} = 5$. This will at most give another factor $(1.5)^\pm$
which depend on $\mu$.  \\
(iii)$T_1 \geq GeV$

Use eq.(\ref{alpha2}) with $T=1GeV$ and gives
\begin{eqnarray}
sin^22\theta < 1.96453 \times 10^{-9}
\end{eqnarray}
Note that we have omitted the change of the $\delta m^2$ due to the
matter effect which will only increase the $\delta m^2$
 as our z is always negative. Increasing $\delta m^2$ will increase the
frequency the the oscillation term $ sin^2( 66.07y)$ 
 and make it easier to be set to 1.

\section*{Acknowledgements}

This work is supported by the National Science Foundation 
Grant No. PHY-0099544.


\begin{thebibliography}{99}

\bibitem{arkani}
N. Arkani-Hamed, S. Dimopoulos and  G. Dvali,
\pl {\bf B429}, 263 (1998); \prd {\bf 59}, 086004 (1999);
I. Antoniadis, N. Arkani-Hamed, S. Dimopoulos and G. Dvali, Nucl. Phys.
{\bf B516}, 70 (1998).

\bibitem{adel} C. D. Hoyle, U. Schmidt, B. R. Heckel, E. G. Adelberger,
J. Gundlach, D. J. Kapner and H. Swanson, hep-ph/0011014.

\bibitem{seesaw}  M. Gell-Mann, P. Ramond and R. Slansky, in {\it
Supergravity}, eds. P. van Niewenhuizen and D.Z. Freedman (North
Holland 1979); T. Yanagida, in Proceedings of {\it Workshop on
 Unified Theory and Baryon number in the Universe}, eds.
O. Sawada and A. Sugamoto (KEK 1979);  R. N. Mohapatra and
G. Senjanovi{\'c}, Phys. Rev. Lett. {\bf 44}, 912 (1980).

\bibitem{hanestad} S. Hanestad, astro-ph/0102290; M. J. Fairbarn,
hep-ph/0101131; L. Hall and D. Smith, Phys. Rev. {\bf D 60}, 085008
(1999).

\bibitem{sn} V. Barger, T. Han, C. Kao and R. J. Zhang, Phys. Lett. {\bf B
461}, 34 (1999);
S. Cullen and M. Perelstein, Phys. Rev. Lett. {\bf 83}, 268 (1999);
C. Hanhardt, J. Pons, D. Phillips and S. Reddy, astro-ph/0102063;
G. Raffelt and S. Hannestad, hep-ph/0103201.



\bibitem{dienes}
K.R. Dienes, E. Dudas and  T. Gherghetta,
Nucl. Phys. {\bf B557}, 25 (1999);
N. Arkani-Hamed, S. Dimopoulos, G. Dvali and J. March-Russell,
hep-ph/9811448.

\bibitem{nandi} R. N. Mohapatra, S. Nandi and A. Perez-Lorenzana,
Phys. Lett. {\bf B466}, 115 (1999); R. N. Mohapatra and
A. Perez-Lorenzana, Nucl. Phys. {\bf B 576}, 466 (2000).

\bibitem{cald} D. O. Caldwell, R. N. Mohapatra and S. Yellin,
Phys. Rev. Lett. {\bf 87}, 041601 (2001); Phys. Rev. {\bf D 64} 
073001 (2001).

\bibitem{das} A. Farragi and M. Pospelov, Phys. Lett. {\bf B 458}, 237
(1999);
G. Dvali and A.Yu. Smirnov, Nucl. Phys. {\bf B563}, 63 (1999). 
A. Ionissian and A. Pilaftsis, hep-ph/9907522; 
C. S. Lam and J. N. Ng, hep-ph/0104129;
G. Mcglaughlin and J. N. Ng, Phys. Lett. {\bf B 470}, 157 (1999);
N. Cosme et al. Phys. Rev. {\bf D 63}, 113018 (2001);
A. Nicolidis and D. T. papadamou, hep-ph/0109048;
C. S. Lam, hep-ph/0110142;
For a review, see C. S. Lam, hep-ph/0108198.

\bibitem{barbieri1}  R. Barbieri, P. Creminelli and A. Strumia,
Nucl. Phys. {\bf B 585}, 28 (2000).

\bibitem{fuller} K. Abazajian, G. M. Fuller, and M. Patel,
hep-ph/0011048.

\bibitem{lukas} A. Lukas, P. Ramond, A. Romanino and G. Ross,
Phys. Lett. {\bf B 495}, 136 (2000);


\bibitem{olive} K. Olive, G. Steigman and T. Walker, Phys. Rep. {\bf 333},
389 (2000); S. Sarkar, Rep. Prog. Phys. {\bf 59}, 1493 (1996).

\bibitem{dolgov} For recent review and references, see A. Dolgov,
hep-ph/0109155; D. Kirilova and M. Chizov, astro-ph/0108341.



\bibitem{fargion} D. Fargion and M. Shepkin, Phys. Lett. {\bf 146B}, 46
(1994).

\bibitem{notzold} D. Notzold and G. Raffelt, Nucl. Phys. {\bf B 307}, 924
(1988).

\bibitem{Dodelson} S.Dodelson and L.M. Widrow, Phys. Rev.  Lett. {\bf 72}, 17
(1984).

\bibitem{barbieri} P. Langacker, S. Petcov, G. Steigman and S. Toshev,
Nucl. Phys. {\bf B 282}, 589 (1987);
R. Barbieri and A. Dolgov, Phys. Lett. {\bf B 237}, 440 (1990);
K. Enquist, K. Kainulainen and J. Maalampi, Phys. Lett. {\bf B 244}, 186
(1990); {\it ibid} {\bf B 249}, 531 (1990); J. Cline,
Phys. Rev. Lett. {\bf
68}, 3137 (1992); D. P. Kirilova and M. Chizov, Phys. Rev. {\bf D 58},
073004 (1998).


\bibitem{FootLep} R.Foot,and R.R.Volkas, Phys. Rev. {\bf D 55},5147 (1997).

 
\end{thebibliography}
\end{document}